\documentclass[%
 reprint,
 amsmath,amssymb,
 aps,
]{revtex4-2}

\usepackage{graphicx}
\usepackage{dcolumn}
\usepackage{bm}
\usepackage{braket}
\usepackage{amsmath}
\usepackage{amssymb}
\usepackage[hidelinks]{hyperref}
\usepackage[capitalize]{cleveref}
\usepackage{empheq}
\usepackage{tikz}
\usepackage{pgfplots}
\pgfplotsset{compat=1.16}
\usepackage{float}
\usepackage{tikz}
\usepackage{subcaption}
\usepackage{array}
\usetikzlibrary{decorations.markings}
\usetikzlibrary{backgrounds,automata}
\usepackage{comment}

\newcommand{\of}[1]{\left( #1 \right)}
\newcommand{\SUM}[2]{\overset{#2}{\underset{#1}{\sum}}}
\newcommand{\p}{\bold{p}}
\newcommand{\K}{\bold{K}}
\newcommand{\Q}{\bold{Q}}
\newcommand{\eqtxt}[2]{\textrm{#1 }\textrm{#2}}

\allowdisplaybreaks
\begin{document}

\preprint{APS/123-QED}

\title{Colossal Magnon Magnetoresistance of Two-Dimensional Magnets}

\author{Caleb Webb}
\author{Shufeng Zhang}%
\affiliation{%
 Department of Physics, University of Arizona\\
}%

\date{\today}

\begin{abstract}
The magnon current holds substantial importance in facilitating the transfer of angular
momentum in spin-based electronics. However, the magnon current in three-dimensional
magnetic materials remains orders of magnitude too small for applications. In contrast, magnon
numbers in two-dimensional systems exhibit significant enhancement and are markedly
influenced by external magnetic fields. Here, we investigate the magnon current in a two-
dimensional easy-axis ferromagnet and find a colossal magnon magnetoresistance (CMMR)
effect, wherein the change of the magnon conductance can reach as high as a thousand
percent in a moderate magnetic field. Moreover, the magnitude of the CMMR exhibits significant
dependence on the orientation of the magnetic field due to the interplay between magnon-
conserving and non-conserving scattering. We propose a non-local magnon-mediated electrical
drag experiment for the possible experimental observation of the predicted effect. With the CMMR effect and a much larger magnon number, magnon current in 2D materials shows promise as a primary source for spin transport in spintronics devices.

\end{abstract}

\maketitle

\section{Introduction}
Magnons, also known as spin wave quanta, are low-energy excitations of magnetic ordered states, holding significance in spintronics for their role in spin transport properties. Analogous to electron spins, magnons carry angular momenta and magnon current serves as a spin current that traverses magnetic media. Transport properties of magnons in various magnetic ordered states, including ferromagnetic, antiferromagnetic, and non-collinear magnetic structures, have been extensively investigated both experimentally and theoretically \cite{1,2,3,4,5,6,7,8,9}. In these studies, the magnon current of
three-dimensional magnets is generated through spin injection from boundaries or a thermal
gradient, however, its magnitude remains orders of magnitude smaller than the conventional
electron spin current, limiting the application of the magnon current as a primary source of spin
information carriers.

Recent advances in experimental development of two-dimensional (2D) magnetic materials \cite{10,11,12,13,14,15,16,17,18,19,20,21,22} provide further opportunity to explore novel magnon transport. In particular, it has been demonstrated \cite{5} that the magnon conductivity in thin YIG films greatly surpasses that of their three dimensional counterparts. In this article we argue that magnon transport should see an even greater enhancement in a truly two-dimensional ferromagnetic insulator. Compared to conventional magnon transport properties in 3D systems, low dimensional magnets experience much stronger spin fluctuations. Owing to the Mermin-Wagner theorem, thermal fluctuations destroy long-range magnetic ordering for the isotropic short-range Heisenberg Hamiltonian \cite{23} and the spatial random field breaks the uniform magnetization into domains \cite{24}. Breaking the continuous $SO(3)$ symmetry by introducing magnetocrystaline anisotropy and/or external field induces a gap, $\Delta$, in the excitation spectrum. In the long-wavelength limit, the magnon dispersion is given by $\omega_k\sim\Delta+Jk^2$ where $J$ is the exchange constant. The equilibrium magnon number $N_0=\int d^2k\left[\exp(\beta\omega_k)-1\right]^{-1}$ (with inverse temperature $\beta$) varies weakly with the gap but diverges logarithmically as $\Delta\rightarrow0$. Since the gap can be controlled by the external magnetic field, one would expect that the magnon conductance $\sigma\sim N_0\tau$, with $\tau$ the magnon lifetime, varies strongly with the external field, implying a larger magnon magnetoresistance (MMR) effect is possible.

In this article we demonstrate that the influence of the external field on the magnon lifetime and gap leads to a colossal MMR in two-dimensinal ferromagnets with easy-axis exhange anisotropy. We find that for fields parallel to the easy-axis the magnon transport properties are governed primarily by changes in equilibrium magnon number. However, non-parallel fields generate non-magnon conserving scattering processes \cite{Chubukov} which greatly reduce the lifetime at small field strengths in 2D. By explicitly calculating the magnon conductance for this model, we find the magnetoresistance could reach thousands of percent for a moderate strength of the magnetic field. Furthermore, MMR is highly dependent on the direction of the field due to the competition between the magnon-conserving and magnon-non-conserving scattering in the model Hamiltonian. Such colossal magnon magnetoresistance (CMMR) is orders of magnitude larger than the conventional magnetoresistance of magnetic materials and is comparable to the magnetic oxides near the metal-insulator transition temperatures \cite{Colossal1,Colossal2}.

The remainder of this article is organized as follows. In section two we introduce the model Hamiltonian and derive the scattering vertices. In section three the scattering widths are discussed. The magnon magneto resistance effect is presented in section four, and a possible experimental realization is discussed in section five.

Throughout this article we set $\hbar=k_B=c=1$.

\section{Model}
\begin{figure}[tph] 
\centering
\begin{tikzpicture}
\draw[->,thick] (0,0) -- (3,0) node[below] {$x$};
\draw[->,thick] (0,0) -- (0,3) node[right] {$z$};
\draw[->,red] (0,0) -- (45:3) node[right] {$\bold{H}_{ext}$};
\draw[->,blue] (0,0) -- (55:3) node[right] {$\bold{M}$} ;
\draw[red] (0,1) arc (90:45:1);
\node[red] at (0.45,1.2) {$\theta_H$};
\draw[blue] (0,0.5) arc (90:55:0.5) node[above] {$\theta$};
\end{tikzpicture}
\caption{Equilibrium magnetization direction $\bold{M}$ in the presence of an external field $\bold{H}_{ext}$ in the $xz$-plane.} \label{mag}
\end{figure}
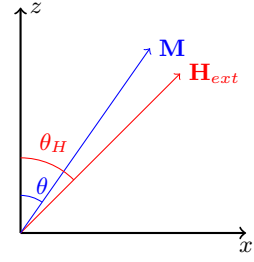

We consider an easy-axis ferromagnetic Heisenberg model on a two-dimensional lattice with out of plane exchange anisotropy and arbitrary external field $\bold{H}_{ext}$, which points at an angle $\theta_H$ from the $z$-axis. The Hamiltonian is
\begin{align} \label{model}
H=-\SUM{\braket{i,j}}{}\of{J\bold{S}_i\cdot\bold{S}_j+J_zS_i^zS_j^z}-\SUM{i}{}\bold{H}_0\cdot\bold{S}_i,
\end{align}
with $J>0$ the isotropic exchange coupling and $J_z>0$ the exchange anisotropy, taken along the positive $z-$axis. $\bold{S}_i$ is the spin operator for lattice site $i$, and in the first term we sum over nearest neighbors $i$, and $j$. The parameter $\bold{H}_0=\mu\bold{H}_{ext}$, with $\mu$ the magnetic moment per unit spin, has been introduced as shorthand. Competition between the anisotropy and external field will shift the ground state magnetization away from the easy axis ($z-$axis). We describe the equilibrium magnetization $\bold{M}$ by an angle $\theta$ as in \cref{mag}.
In order to expand \cref{model} about its classical ground state, the spin operators are first expressed in a rotated frame, $\bold{S}=\bold{R}\bold{S'}$, with $z'$-axis parallel to $\bold{M}$. In this rotated frame \cref{model} becomes
\begin{widetext}
\begin{equation} \label{rotatedmodel}
H=\begin{aligned}[t] -\SUM{\braket{i,j}}{}&J\bold{S'}_i\cdot\bold{S'}_j+J_z\left[\sin^2\theta {S'}_i^x{S'}_j^x+\cos^2\theta {S'}_i^z{S'}_j^z-\sin\theta\cos\theta\of{{S'}_i^x{S'}_j^z+{S'}_i^z{S'}_j^x}\right]-\\
&-\SUM{i}{}H_0^x\of{\cos\theta {S'}_i^x+\sin\theta {S'}_i^z}+H_0^z\of{-\sin\theta{S'}_i^x+\cos\theta{S'}_i^z},\end{aligned}
\end{equation}
\end{widetext}
which may now be expanded in small fluctuations by means of the usual Holstein-Primakoff (HP) transformation. Up to cubic order in the boson operators we have
\begin{align} \label{HP}
\begin{aligned} &{S'}_i^z=S-a_i^{\dagger}a_i \\
&{S'}_i^{+}=\sqrt{2S}\of{1-\frac{a_i^{\dagger}a_i}{4S}}a_i\\
&{S'}_i^{-}=\sqrt{2S}a_i^{\dagger}\of{1-\frac{a_i^{\dagger}a_i}{4S}}.
\end{aligned}
\end{align}
This truncation of the HP transformation is sufficient to produce the lowest order corrections to the magnon lifetime at low temperatures ($T<<T_C$ with $T_C$ the Curie temperature), where the magnon density is not too large. Applying \cref{HP} to \cref{rotatedmodel} yields the Hamiltonian for the interacting magnon system, $H=H^{(0)}+H^{(1)}+H^{(2)}+H^{(3)}+H^{(4)}$ with $H^{(n)}$ containing $n$ boson operators.  The first two terms are
\begin{align}
&H^{(0)}=-SN\of{\alpha-\alpha_z\cos^2\theta-H_0\cos(\theta_H-\theta)}\\
&H^{(1)}=\sqrt{\frac{S}{2}}\of{\alpha_z\sin2\theta-H_0\sin(\theta_H-\theta)}\SUM{i}{}a_i^{\dagger}+a_i
\end{align}
and describe the classical ground state energy and shift due to uniform canting of the lattice spins towards the anisotropy axis, respectively. Above, we have defined the anisotropy and exchange parameters $\alpha_z=J_zSN_{\delta}$ and $\alpha=JSN_{\delta}$, with $N_{\delta}$ the number of nearest neighbors for each lattice site. $N$ is the total number of lattice sites. For simplicity, we assume a lattice with inversion symmetry. The angle $\theta$ is defined by minimizing $H^{(0)}$, or equivalently, setting $H^{(1)}=0$, and is given by the solution to
\begin{align} \label{theta}
\alpha_z\sin(2\theta)=H_0\sin(\theta_H-\theta).
\end{align}

The interacting magnon Hamiltonian is then obtained by application of the Fourier transform $a_i=\frac{1}{\sqrt{N}}\SUM{k}{}e^{i\bold{k}\cdot\bold{R}_i}a_k$
\begin{align}
H=&\SUM{k}{}\omega_ka_k^{\dagger}a_k+V^{(2)}_{k}\of{a_ka_{-k}+a^{\dagger}_{-k}a^{\dagger}_{k}}\nonumber\\
+&\frac{1}{\sqrt{N}}\SUM{k_1,k_2,k_3}{}V^{(3)}_{k_1,k_2,k_3}\of{a^{\dagger}_{k_1}a^{\dagger}_{k_2}a_{k_3}+a^{\dagger}_{k_3}a_{k_2}a_{k_1}}\nonumber\\
+&\frac{1}{N}\SUM{k_1,k_2,k_3,k_4}{}V^{(4)}_{k_1,k_2,k_3,k_4}a^{\dagger}_{k_1}a^{\dagger}_{k_2}a_{k_3}a_{k_4}. \label{magnonH}
\end{align}

Defining the lattice factor $\gamma_k=\frac{1}{N_{\delta}}\SUM{\delta}{}e^{-\bold{k}\cdot\bold{\delta}}$, with $\bold{\delta}$ a vector from each site to its nearest neighbors, the full magnon dispersion and interaction vertices are
\begin{align}
&\omega_k=-(2\alpha+\alpha_z\sin^2\theta)\gamma_k+2(\alpha+\alpha_z\cos^2\theta)+H_0\cos(\theta_H-\theta)\\
&V^{(2)}_k=-\frac{\alpha_z\sin^2(\theta)}{2}\gamma_k\\
&V^{(3)}_{k_1,k_2,k_3}=-\frac{\delta_{k_1+k_2-k_3}}{\sqrt{2S}}\alpha_z\sin(2\theta)\gamma_{k_1}\label{v3}\\
&V^{(4)}_{k_1,k_2,k_3,k_4}=\frac{\delta_{k_1+k_2-k_3-k_4}}{2S}\alpha\of{\gamma_{k_4}+\gamma_{k_1}-2\gamma_{k_2-k_3}}.\label{v4}
\end{align}
Since we have limited our analysis to the lowest order corrections to the magnon lifetime, it is sufficient to consider only terms up to order $\mathcal{O}(a^4)$. Moreover, the off diagonal terms $V^{(2)}_{k}\of{a_ka_{-k}+a^{\dagger}_ka^{\dagger}_{-k}}$ are treated as scattering processes in the Dyson series for the self energy. To lowest order in $\alpha_z$ they do not contribute any broadening and are from here on neglected. Our treatment is similar to that in \cite{Chubukov2} in the limit $T>>\alpha_z$. Additionally, in $H^{(4)}$ we have neglected terms proportional to $\alpha_z$ on the grounds that $J_z<<J$, and so the two-magnon scattering processes will be dominated by the exchange coupling.

For a square lattice in the long wavelength limit the magnon frequency reduces to the quadratic form $\omega_k=\Delta+\omega_0k^2$ with gap
\begin{align} \label{gap}
    \Delta=\alpha_z(3\cos^2\theta-1)+H_0\cos(\theta_H-\theta)
\end{align}
and frequency
\begin{align}
    \omega_0=\frac{2\alpha+\alpha_z\sin^2\theta}{4}\approx \alpha/2.
\end{align}
The frequency is governed primarily by the exchange coupling, and is largely unaffected by the external field. The gap, \cref{gap}, is strongly dependent on the field strength and has a somewhat complicated relationship with the field direction. Numerically analysis of \cref{gap,theta} shows that $\Delta$ develops a minimum for $H_0>0$ when $\theta_H\geq\theta_c\approx0.424\pi$. This minimum reaches its lowest value of $\Delta_{min}=\alpha_z$ when $\theta_H=\pi/2$ and at the critical field $H_{c}=2\alpha_z$. For $\theta_H<\theta_c$ the gap is a monotonically increasing function of the external field.

In this limit the interaction vertices reduce to
\begin{align}
&V^{(3)}_{k_1,k_2,k_3}=-\frac{\delta_{k_1+k_2-k_3}}{\sqrt{2S}}\alpha_z\sin(2\theta)\of{1-k_1^2/4}\label{V3}\\
&V^{(4)}_{k_1,k_2,k_3,k_4}=\frac{\delta_{k_1+k_2-k_3-k_4}}{8S}\alpha\of{k_4^2+k_1^2-2\of{k_2-k_3}^2}.\label{V4}
\end{align}

The potential $V^{(3)}$ in \cref{magnonH} describes the spontaneous decay of one magnon into two, and vice versa, and is strongly dependent on the direction of the external field. From \cref{theta} one finds that for $\theta_H=0$, $\theta=0$ and for field strengths $H_0>2\alpha_z$, $\theta_H=\pi/2\implies\theta=\pi/2$. We conclude that non-magnon conserving processes ($V^{(3)}$) vanish for external fields parallel to the easy axis, or for fields $H_0>H_c=2\alpha_z$ perpendicular to it. The final term in \Cref{magnonH} describes magnon-magnon scattering with potential $V^{(4)}$ which is independent of the equilibrium magnetization direction (see \cref{V4}). Importantly, this potential is directly proportional to the magnon energy, $V^{(4)}\sim J k^2$, which is assumed small. For thermal magnons $Jk^2\sim T$, and these two vertices may be comparable over a wide range of magnon energy for temperature well below $T_C$.

\section{Scattering Widths}
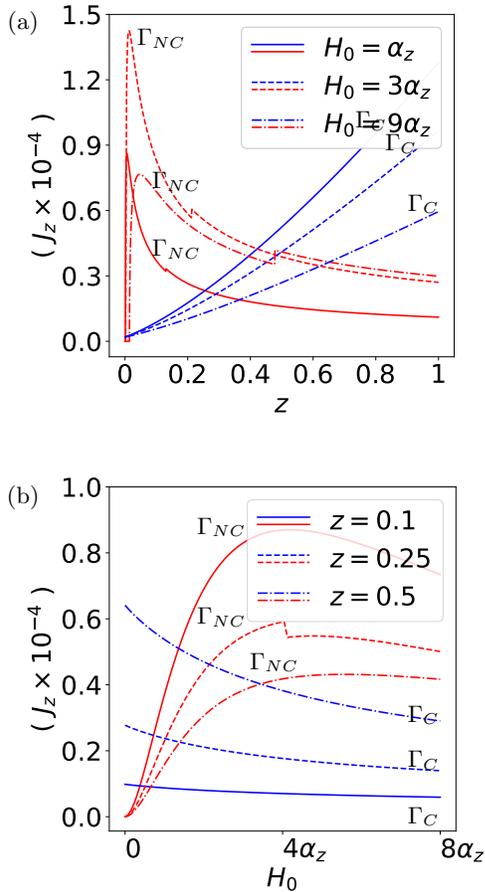
\begin{figure}[tph]
\centering
\begin{tikzpicture}
\node at (-0.1,0) {\includegraphics[width=0.327\textwidth]{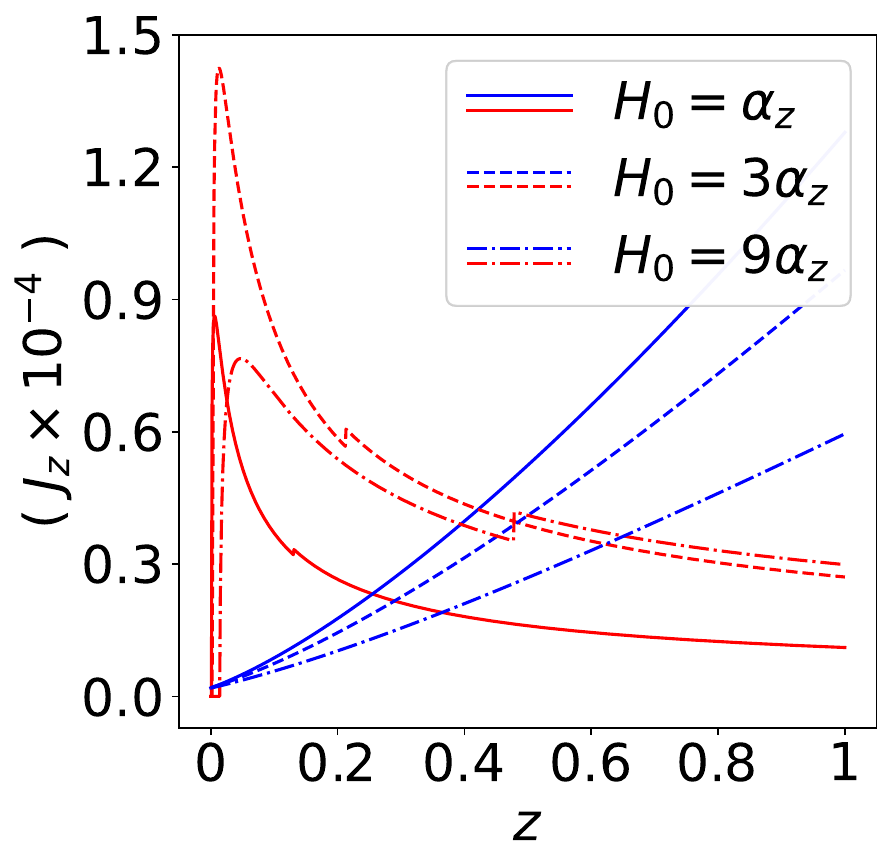}};
\node[anchor=west] at (-1.4,-0.5) {$\Gamma_{NC}$};
\node[anchor=west] at (-1.4,0.4) {$\Gamma_{NC}$};
\node[anchor=west] at (-1.6,2.3) {$\Gamma_{NC}$};
\node[anchor=west] at (1.3,1.2) {$\Gamma_{C}$};
\node[anchor=west] at (1.7,0.9) {$\Gamma_{C}$};
\node[anchor=west] at (2.0,0.1) {$\Gamma_{C}$};
\node at (-3,2.5) {(a)};
\node at (0.1,-6.3) {\includegraphics[width=0.35\textwidth]{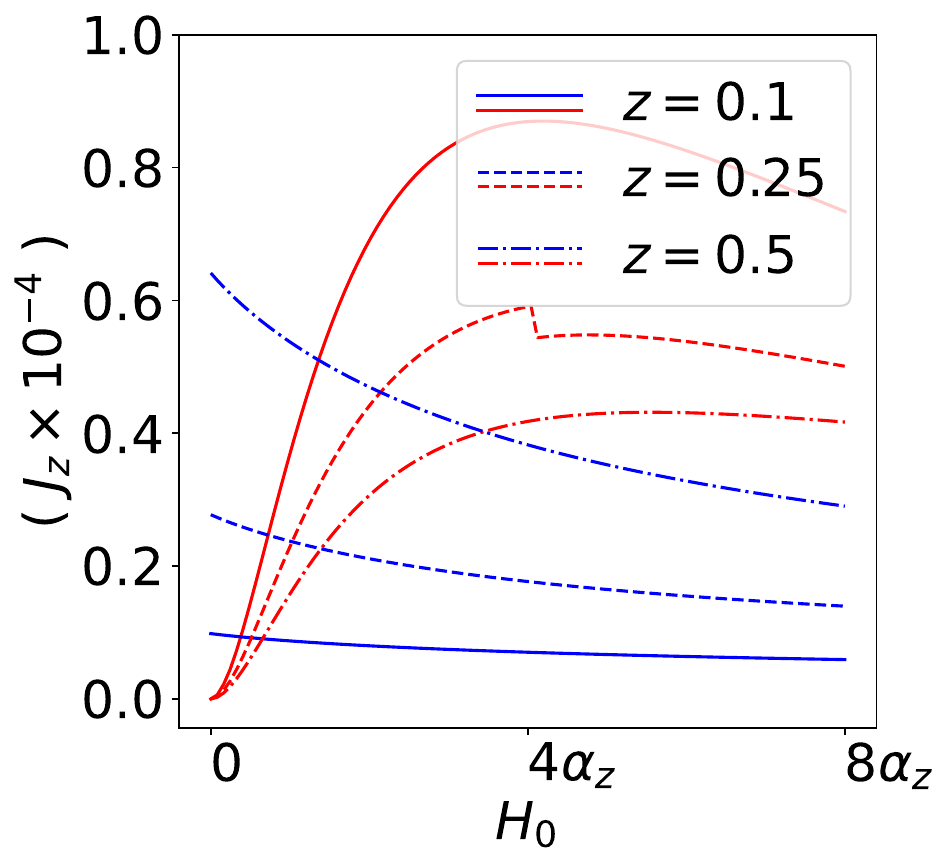}};

\node[anchor=west] at (-0.8,-4.2) {$\Gamma_{NC}$};
\node[anchor=west] at (-0.8,-5.4) {$\Gamma_{NC}$};
\node[anchor=west] at (-0.1,-6) {$\Gamma_{NC}$};

\node[anchor=west] at (2.0,-6.7) {$\Gamma_{C}$};
\node[anchor=west] at (2.0,-7.3) {$\Gamma_{C}$};
\node[anchor=west] at (2.0,-8.05) {$\Gamma_{C}$};
\node at (-3,-3.8) {(b)};

\end{tikzpicture}

\caption{\label{gammacompare}(Color Online) Comparison of the non-magnon conserving 
 and magnon conserving scattering widths for external field at an angle $\theta_H=\pi/6$ to the easy-axis. (a) Scattering widths as a function of the dimensionless energy parameter $z=\omega_0k^2/T$ for different external field strengths $H_0$. (b) Widths as a function of external field strength for different $z$ in the range $0<z<1$. We use red curves for $\Gamma_{NC}$ and blue for $\Gamma_C$. The parameters chosen here are $J_z/J=10^{-5}$ and $T/T_C=10^{-2}$. We assume a square lattice in the long-wavelength limit, for which the Curie temperature is approximately $T_C/J\sim0.1$. The momentum cutoff $z_c$ is chosen so that $\omega_k\lesssim T$.}
\end{figure}

The scattering widths for each process are calculated to single-loop order using the standard diagramatic techniques \cite{Mahan,Baym}. We denote by $\Gamma_{NC}$ the width for non-magnon conserving scattering ($V^{(3)}$), and by $\Gamma_C$ for magnon conserving scattering ($V^{(4)}$).These scattering widths are plotted as a function of the dimensionless energy parameter $z=\omega_0k^2/T$ for several different external field magnitudes in \cref{gammacompare}.
\noindent The results presented above are for the field angle $\theta_H=\pi/6$, but remain qualitatively similar for $0<\theta_H<\theta_c$. For parallel fields, $\Gamma_{NC}=0$ but otherwise the results for $\Gamma_C$ are mostly unaffected. Fields at angles $\theta_H\geq\theta_c$ have a more significant influence on $\Gamma_{NC}$, and this case is discussed separately below. Full details of these calculations can be found in the appendix. Here, we briefly summarize some of the more salient features found within \Cref{gammacompare}.

From Fig. 2a it is apparent that the width $\Gamma_C$ is suppressed by an increasing external field, consistent with the notion that increasing the gap suppresses the number of magnons- thereby reducing the scattering width. The relationship between $\Gamma_{NC}$ and $H_0$ is, however, more complex. This is shown more clearly in Fig. 2b, in which we plot each scattering widths as a function of external field strength at several values of $z$ within the available momentum range. The peak in $\Gamma_{NC}$ results from competition between the strengthening vertex (\cref{V3}) and widening gap. As $H_0$ increases, the magnetization angle $\theta$ increases from zero and approaches the limit $\theta_H$. At the same time, the gap (\cref{gap}) increases monotonically for $\theta_H<\theta_c$. The width $\Gamma_{NC}$, then, increases from zero until the magnetization angle is saturated by the external field- at which point the increase in vertex strength is overtaken by the growth in $\Delta$.

\begin{figure}[tph]
\centering
\begin{subfigure}{0.45\columnwidth}
\centering
\subcaption{}\includegraphics[width=\textwidth]{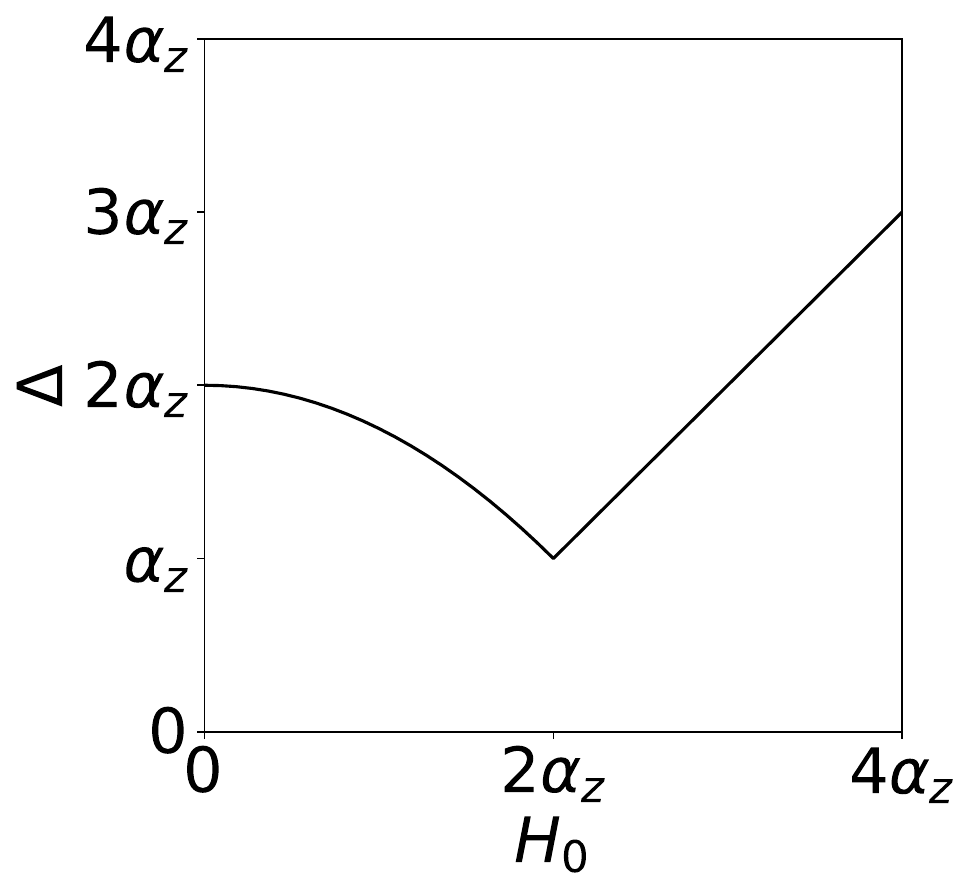}
\end{subfigure}
\hspace{0.1cm}
\begin{subfigure}{0.45\columnwidth}
\centering
\subcaption{}\includegraphics[width=\textwidth]{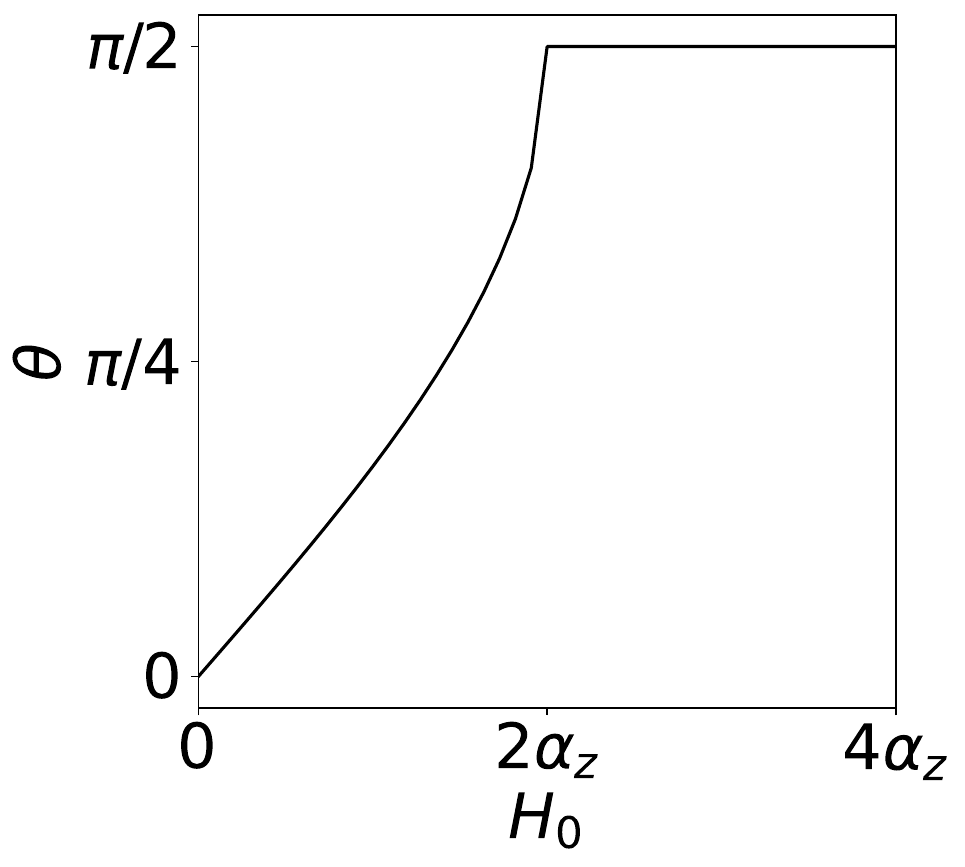}
\end{subfigure}

\hspace{0.1cm}
\begin{subfigure}{0.45\columnwidth}
\centering
\subcaption{}\includegraphics[width=\textwidth]{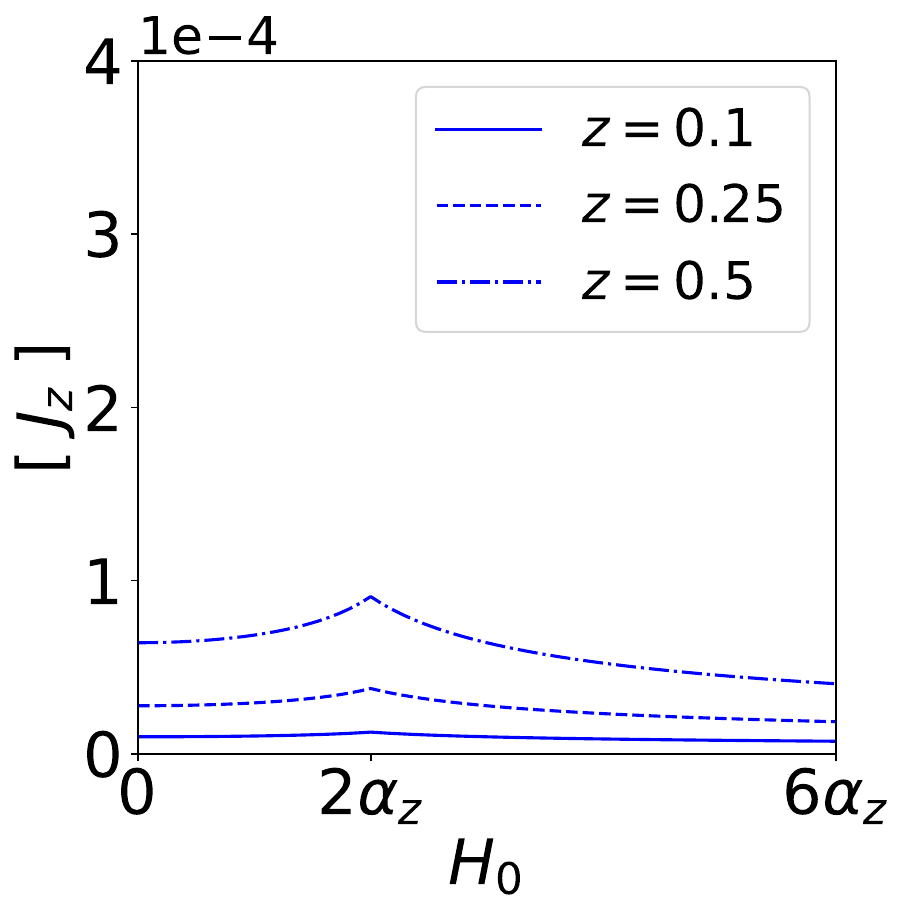}
\end{subfigure}
\hspace{0.1cm}
\begin{subfigure}{0.45\columnwidth}
\centering
\subcaption{}\includegraphics[width=\textwidth]{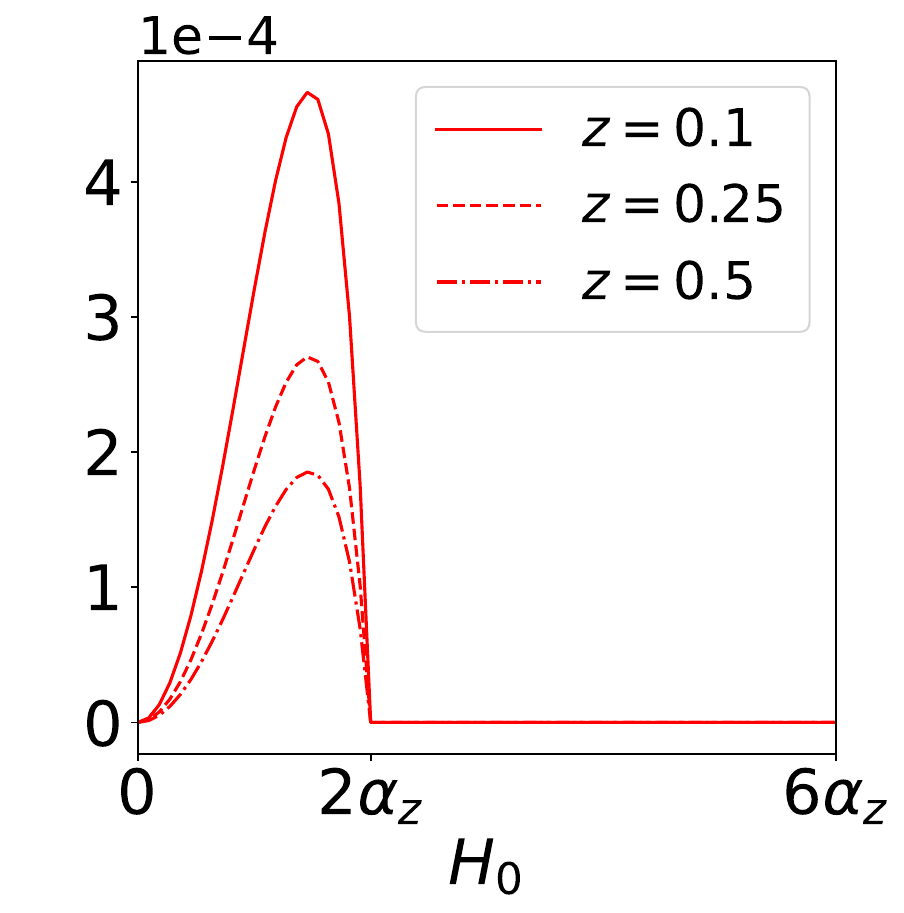}
\end{subfigure}
\caption{\label{Disc} (a) Energy gap, (b) Equilibrium magnetization direction, (c) Magnon-conserving scattering width, and (d) Non-magnon conserving scattering width as a function of external field strength for in-plane fields. All other parameters are as in \cref{gammacompare}.}
\end{figure}

For field angles $\theta_H\geq\theta_c$ the gap does not increase monotonically with field, but develops a minimum for small $H_0$. This is most pronounced for in-plane fields, where the minimum is reached exactly as the magnetization angle is saturated by the external field, $\theta\rightarrow\pi/2$. Both scattering widths should therefore be \textit{increasing} for $0\leq H_0<2\alpha_z$. The vertex $V^{(3)}$, however, vanishes when the magnetization switches to in-plane, and so $\Gamma_{NC}$ drops abruptly to zero from its maximum value at the critical field $H_c$. The full scattering width, then, will change dramatically as the field is tuned across this critical point (see \cref{Disc}).

\section{Magnon Magnetoresistance}
The magnon conductivity can be calculated from a linearized Boltzmann equation in the relaxation time approximation $\sigma=\int d^2p v^2 \partial n/\partial \omega \tau$. Or,
\begin{align} \label{sigma}
\sigma=\frac{4\pi}{\beta}\int_0^{z_{c}}dz\frac{z e^{z+\beta\Delta}}{\Gamma(z)\of{e^{z+\beta\Delta}-1}^2},
\end{align}
where $\Gamma=\Gamma_{NC}+\Gamma_C$. From \cref{gammacompare,Disc} it is clear that the non-conservative scattering is dominant in the small momentum regime, where the Bose distribution is largest. One therefore anticipates that these processes have a greater influence over the conductivity. Indeed, we find a large MMR effect for non-parallel external fields. Of particular interest are in-plane fields, for which we find a colossal spike in the magnon resistivity at fields $H_0<2\alpha_z$ which drops rapidly to zero at the critical point.
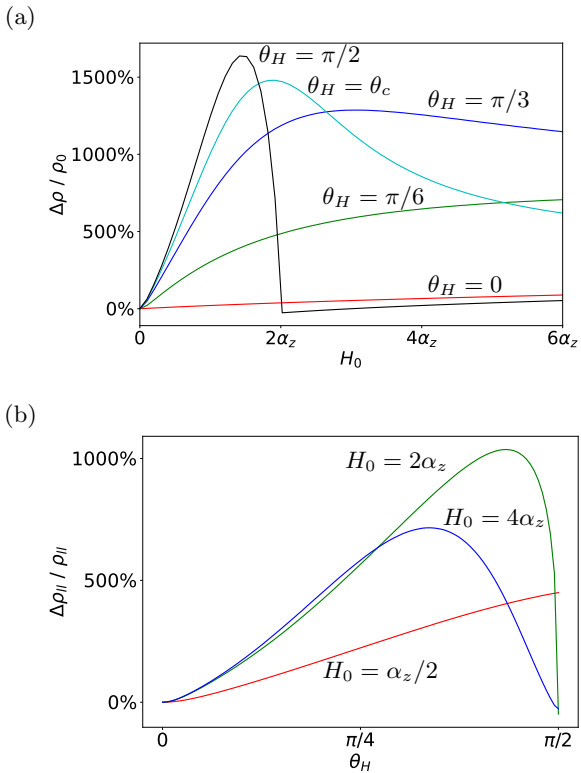
\begin{figure}[tph]
\hspace{-1cm}
\begin{tikzpicture}
\node at (-0.1,0) {\includegraphics[width=0.4\textwidth]{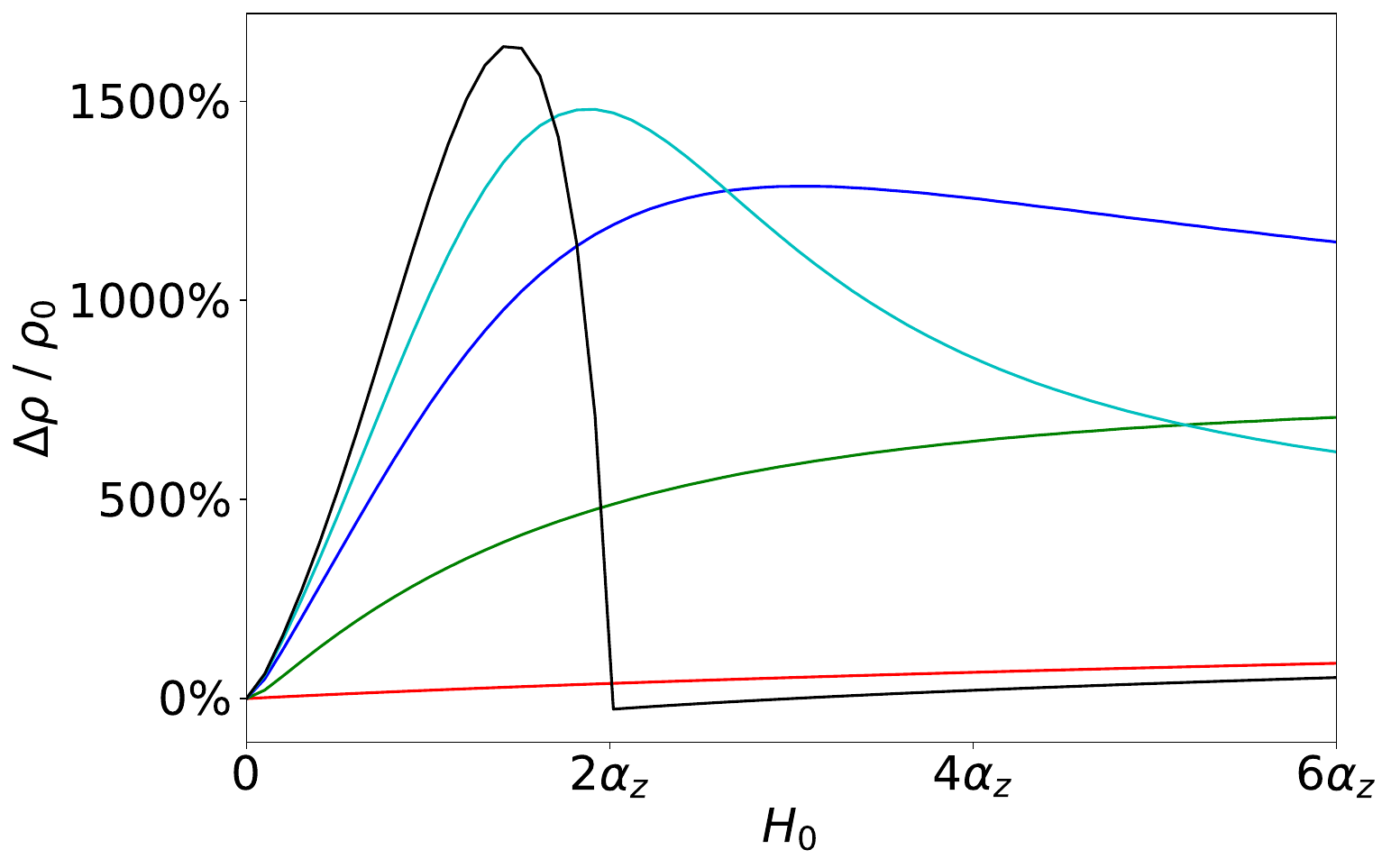}};
\node[anchor=west] at (-0.95,2.0) {$\theta_H=\pi/2$};
\node[anchor=west] at (-0.3,1.6) {$\theta_H=\theta_c$};
\node[anchor=west] at (1.3,1.4) {$\theta_H=\pi/3$};
\node[anchor=west] at (-0.1,0.1) {$\theta_H=\pi/6$};
\node[anchor=west] at (1.3,-1.05) {$\theta_H=0$};

\node at (-4,2.5) {(a)};
\node at (-0.1,-5.3) {\includegraphics[width=0.4\textwidth]{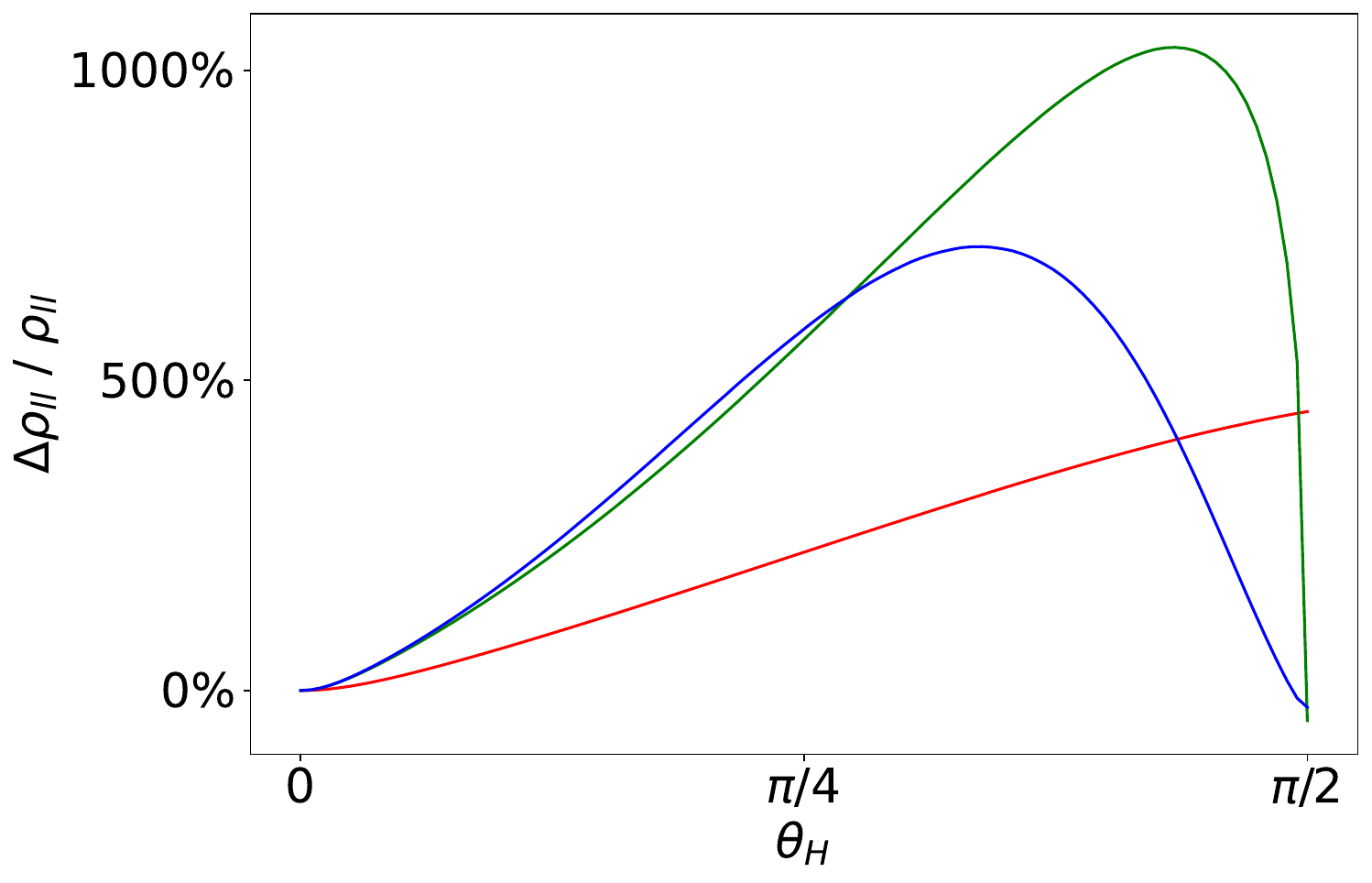}};

\node[anchor=west] at (-0.1,-6.2) {$H_0=\alpha_z/2$};
\node[anchor=west] at (0.2,-3.4) {$H_0=2\alpha_z$};
\node[anchor=west] at (1.5,-4.15) {$H_0=4\alpha_z$};

\node at (-4,-2.8) {(b)};

\end{tikzpicture}
\caption{\label{rho}(Color Online) CMMR effect as a function of external field strength and direction. (a) Magnon magnetoresistance ratio at several different directions of the external field; (b) Anisotropic Magnon magnetoresistance ratio as a several different strengths of the field.}
\end{figure}
In Fig. (4a) we present the magnon magnetoresistance ratio as a function of external field strength. In the figure $\rho_0=\rho(H_0=0)$ and $\Delta\rho=\rho-\rho_0$. The magnon conductivity is also influenced by the direction of the external field, and the anisotropic magnon magnetoresistance ratio is plotted in Fig. (4b). Here $\rho_{||}=\rho(\theta_H=0)$ and $\Delta\rho_{||}=\rho-\rho_{||}$. That this (CMMR) should be so large can, again, be seen as a consequence of the sharply peaked width $\Gamma$ due to non-conservative scattering. Heuristically, $\rho\sim\braket{\Gamma}/N_0$ with $\braket{\cdot}$ denoting the thermal average over the Bose distribution. Due to its logarithmic dependence on the gap, $N_0$ is only slightly influenced by the external field compared to $\Gamma$ for non-parallel fields. The MMR is then governed primarily by the non-conservative scattering processes at low temperatures.

It should be noted that the predicted CMMR is most pronounced for field strengths $H_0\sim \alpha_z$, and is significantly less pronounced for larger fields. The anisotropic MMR, however, remains significant for $H>H_c$ (see Fig. (4b). At the critical field, the magnon resistivity is enhanced by as much as $1000\%$ relative to its value at parallel field. However, even at much larger fields the MMR can reach up to $500\%$. 

\section{Experimental Realization} 
We now elaborate on potential experimental approaches to realize the predicted CMMR effect. Typically, the magnon current is investigated either by the spin Seebeck effect \cite{1} through inducing a temperature gradient or by applying an electric current in a heavy metal, thus injecting a spin Hall current into the contacting magnetic layer \cite{4,5}. Here, we focus on the latter method, wherein the magnetic layer is a 2D magnet insulator. The setup proposed here differs from the magnon conductivity measurements in thin YIG films of \cite{5} only in that the magnetic layer is two-dimensional, and the external field is allowed out of the sample plane.

\begin{figure}[tph]
\centering
\includegraphics[width=0.9\columnwidth]{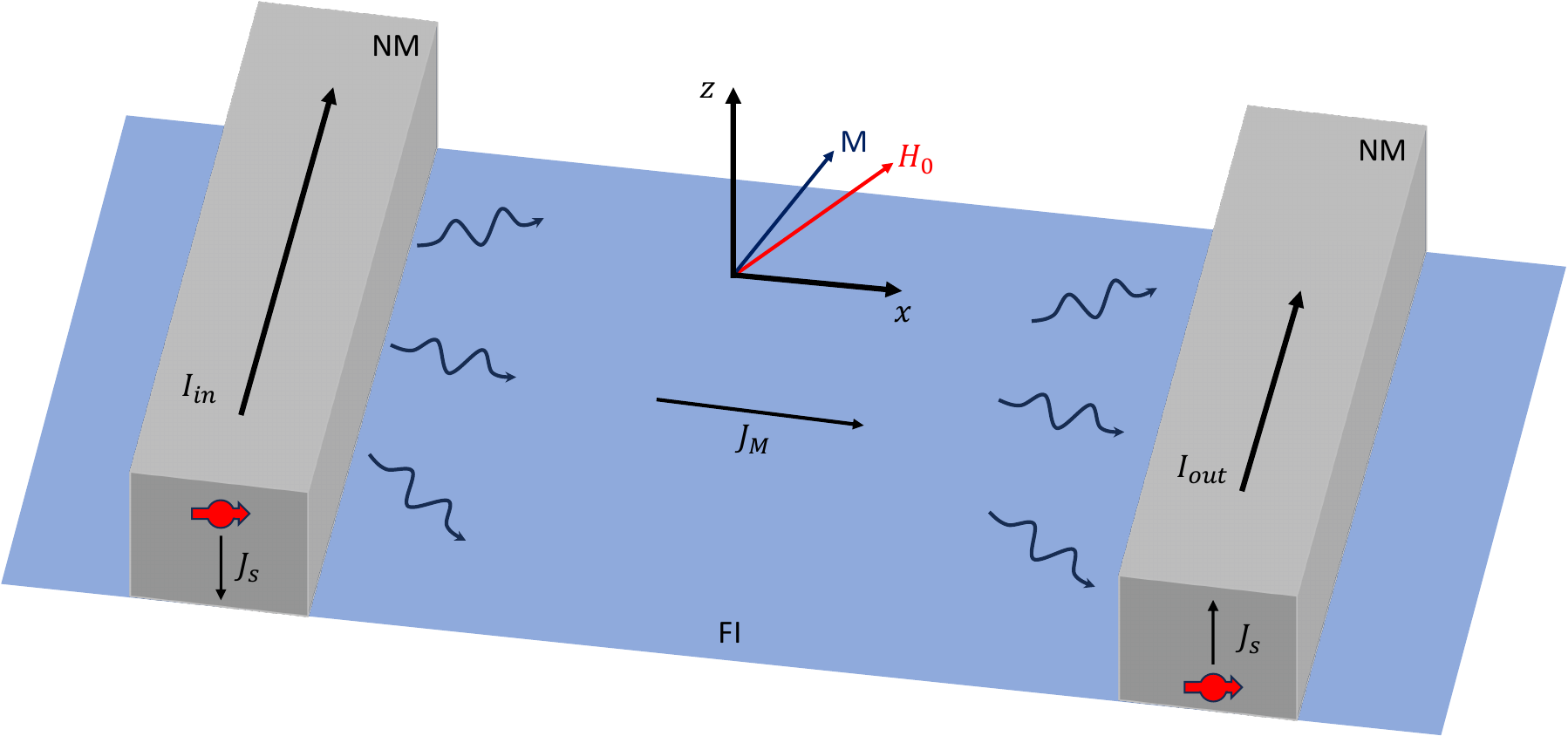}
\caption{\label{scheme}Schematic of experimental setup for non-local electrical drag measurement. Two metalic strips (NM) are placed some distance apart on a two dimensional ferromagnetic insulator (FI). Driving a current ($I_{in}$) in the left strip injects a spin current ($J_s$) into the FI via the spin Hall effect. This in turn excites a magnon current ($J_M$) which is transported diffusively through the FI where it encounters the other NM strip. Here, the magnon current excites a spin current which is then converted to an electrical current $I_{out}$ via the inverse spin Hall effect.}
\end{figure}

As in \cref{scheme} the injected spin current excites a magnon current in the magnetic layer which is then converted into a spin current in another metal layer some distance away. The spin current is then converted to an electrical current via the inverse spin Hall effect. The resistance of the device can be readily determined from the non-local voltage produced, from which the magnon conductivity can be deduced. 

Given that the spin direction of the spin Hall current from the heavy metal is in-plane, the magnon current would predominantly respond to the in-plane component of the magnetization. By manipulating the magnitude and orientation of the applied magnetic field, one can align the magnetization in any desired direction, accommodating either perpendicular or in-plane anisotropy of the 2D magnet. Assuming magnon scattering dominates over other forms of scattering such as magnetic defects, the observed inverse spin Hall signals in the detecting heavy metal bar would delineate the magnitude and directional dependence of the magnetic field, effectively manifesting the predicted CMMR effect.

\appendix
\renewcommand\thefigure{\thesection\arabic{figure}} 
\section{Derivation of scattering widths}
\setcounter{figure}{0} 
In this section we derive in detail the two scattering widths, $\Gamma_{NC}$ and $\Gamma_C$, from the usual Matsubara formalism in Greens function theory.

\subsection{Non-conservative scattering}
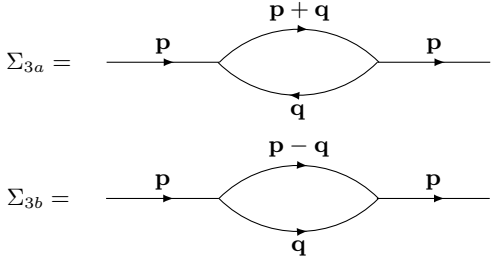
\begin{figure}[tph]
\centering
\begin{subfigure}{\columnwidth}
\centering
\begin{tikzpicture}[scale=1.5]
\node at (-0.6,0) {$\Sigma_{3a}=$};
    \draw[decoration={markings,mark=at position 0.6 with {\arrow{latex}}},postaction=decorate] (0,0) -- (1,0) node[pos=0.5,above] {$\bold{p}$};
    \draw[decoration={markings,mark=at position 0.55 with {\arrow{latex}}},postaction=decorate] (1,0) arc (135:45:1) node[pos=0.5,above] {$\bold{p+q}$};
    \draw[decoration={markings,mark=at position 0.55 with {\arrow{latex}}},postaction=decorate] (2.4,0) arc (-45:-135:1) node[pos=0.5,below] {$\bold{q}$};
    \draw[decoration={markings,mark=at position 0.6 with {\arrow{latex}}},postaction=decorate] (2.4,0) -- (3.4,0) node[pos=0.5,above] {$\bold{p}$};
\end{tikzpicture}
\end{subfigure}
\hspace{0.5cm}
\begin{subfigure}{\columnwidth}
\centering
\begin{tikzpicture}[scale=1.5]
\node at (-0.6,0) {$\Sigma_{3b}=$};
    \draw[decoration={markings,mark=at position 0.6 with {\arrow{latex}}},postaction=decorate] (0,0) -- (1,0) node[pos=0.5,above] {$\bold{p}$};
    \draw[decoration={markings,mark=at position 0.55 with {\arrow{latex}}},postaction=decorate] (1,0) arc (135:45:1) node[pos=0.5,above] {$\bold{p-q}$};
    \draw[decoration={markings,mark=at position 0.55 with {\arrow{latex}}},postaction=decorate] (1,0) arc (-135:-45:1) node[pos=0.5,below] {$\bold{q}$};
    \draw[decoration={markings,mark=at position 0.6 with {\arrow{latex}}},postaction=decorate] (2.4,0) -- (3.4,0) node[pos=0.5,above] {$\bold{p}$};
\end{tikzpicture}
\end{subfigure}
\caption{\label{Sigma3} All single-loop order corrections to the Self-Energy generated by the cubic terms in \cref{magnonH}. Vertices are given by \cref{V3}.}
\end{figure}

Up to one-loop order in the Dyson series, the second line in \cref{magnonH} generates the two self-energy diagrams shown in \cref{Sigma3}. The polarization-bubble type diagram, $\Sigma_{3a}$, is a finite-temperature effect which dominates the magnon lifetime for $T>0$. $\Sigma_{3b}$, on the other hand, describes spontaneous magnon decay and is present even at $T=0$ \cite{chernyshev,villan}.

Following the Matsubara formalism these corrections are calculated as, respectively,
\begin{widetext}
\begin{align}
    \Sigma_{3a}(\omega,\bold{k})&=\frac{1}{N}\SUM{q}{}\left[V^{(3)}_q+V^{(3)}_k\right]^2\frac{n(\omega_{k+q})+n(\omega_{q})}{i\omega-\omega_{k+q}+\omega_q} \label{sig3a} \\
    \nonumber\\
        \Sigma_{3b}(\omega,\bold{k})&=\frac{1}{N}\SUM{q}{}V^{(3)}_{k-q}\left[V^{(3)}_{k-q}+V^{(3)}_q\right]\frac{1+n(\omega_{q})-n(\omega_{k-q})}{i\omega-\omega_{q}-\omega_{k-q}} \label{sig3b} \\
        \nonumber
\end{align}
\end{widetext}
Above, $n(\omega)=\of{e^{\beta\omega}-1}^{-1}$ is the equilibrium magnon distribution at inverse temperature $\beta=1/T$, and all sums over constrained momenta have been carried out so that the vertices read $V^{(3)}_k=V_0\of{1-\frac{1}{6}k^2}$ in the long-wavelength limit, where $V_0=\frac{\alpha_z}{\sqrt{2S}}\sin(2\theta)$. Within this limit integration is commonly carried out over all $\bold{q}\in\mathbb{R}^2$, reasoning that any error introduced by including large $q$ should be suppressed by the Bose distributions. However, these processes involve spontaneous creation/annihilation of magnons. Due to the presence of the gap, $\omega_k=\Delta+\omega_0k^2$, the initial and final states must have a finite energy, and so there must be a lower bound for the external momenta in \cref{Sigma3}. As will be shown below, this lower bound is sensitive to the choice of integration cutoff for the internal momentum, $q_{c}$.

The scattering widths for \cref{sig3a,sig3b} are constrained on-shell by the following kinematic relations. $\Gamma_{3a}(\bold{k})=-2\textrm{Im}\of{\Sigma_{3a}}>0$ only for momenta $\bold{k}$ satisfying $\omega_k=\omega_{k+q}-\omega_q\implies 2\omega_0\bold{k}\cdot\bold{q}=\Delta$. For finite internal momentum $q$, then, the available scattering states are bounded below by $k>\Delta/2\omega_0q_{c}$. $\Gamma_{3b}$ is constrained by $\omega_k=\omega_{k-q}+\omega_q$, which is satisfied only for $q\in\left[q_-,q_+\right]$ with
\begin{align}
    q_{\pm}=\frac{1}{2}\of{k\pm\sqrt{k^2-2\Delta/\omega_0}}.
\end{align}
These decay processes are therefore only allowed for magnons with a kinetic energy $\omega_0k^2>2\Delta$, which is a sensible constraint for the spontaneous emission of two magnons with a gapped spectrum.

The momentum cutoff is chosen so that $\omega_k\lesssim T$, and so the Bose distributions can be roughly approximated as $n(\omega)\sim 1/\beta\omega$. Then
\begin{align}
    \Gamma_{3a}\approx\frac{2V_0^2}{\pi\beta}\int d^2\bold{q}\of{\frac{1}{\omega_q}+\frac{1}{\omega_{p}+\omega_q}}\delta\of{2\omega_0pq\cos\phi-\Delta},
\end{align}
where we have ignored the momentum dependence of the vertex. The $\delta-$function is used to eliminate the integral over angle, yielding
\begin{align}
    \Gamma_{3a}=\frac{2V_0^2q_0}{\pi\beta\Delta}\int_{q_0}^{q_{c}}\frac{qdq}{\sqrt{q^2-q_0^2}}\of{\frac{1}{\omega_q}+\frac{1}{\omega_q+\omega_p}}
\end{align}
with $q_0(k)=\Delta/2\omega_0k$. As $k\rightarrow\frac{\Delta}{2\omega_0q_{c}}$, $\Gamma_{3a}\rightarrow0$. Similarly,
\begin{align}
\begin{aligned}
    \Gamma_{3b}=\frac{V_0^2}{\pi\beta}\int_{q_-}^{q_+}&\frac{qdq}{\sqrt{(q_+^2-q^2)(q^2-q_-^2)}}\times \\ &\times \of{1+\frac{1}{\omega_q}-\frac{1}{\omega_k-\omega_q}}
    \end{aligned}
\end{align}

Both integrations are easily carried out. In terms of the dimensionless variables $z=\beta\omega_0k^2$, $z_{c}=\beta\omega_0q_{c}^2$ and $z_-=\beta\omega_0q_0^2(q_c)$
\begin{widetext}
\begin{align}
    &\Gamma_{3a}=\frac{V_0^2}{\pi\omega_0}\left[\frac{\tan^{-1}\of{\sqrt{\frac{4z_c(z-z_-)}{\beta\Delta(4z+\beta\Delta)}}}}{\sqrt{\beta\Delta(4z+\beta\Delta)}}+\frac{\tan^{-1}\of{\sqrt{\frac{4z_c(z-z_-)}{(2z+\beta\Delta)^2+4z\beta\Delta}}}}{\sqrt{(2z+\beta\Delta)^2+4z\beta\Delta}}\right]
\end{align}
\end{widetext}
To leading order in $\beta\Delta$ this width scales as $\Gamma_{3a}\sim 2V_0^2/\omega_0\sqrt{\beta\Delta}$. This suppression by the gap can be understood as a consequence of the decrease in magnon number. As $\Delta$ increases, there are fewer magnons in the system, and therefore fewer scattering events.

For the spontaneous decay we obtain
\begin{align}
    \Gamma_{3b}=\frac{V_0^2}{8\omega_0}\Theta\of{z-2\beta\Delta}.
\end{align}
Apart from the constraint $z>2\beta\Delta$ this width is independent of momentum, temperature and gap. For small fields $\Gamma_{3a}>>\Gamma_{3b}$ and so this spontaneous decay process has only a small influence on the magnon lifetime above zero temperature. The discontinuity at $z=2\beta\Delta$ is responsible for the sudden jumps in $\Gamma_{NC}$ in \cref{gammacompare}.

\subsection{Conservative scattering}
Following the general procedure for calculating self-energy corrections due to two-particle scattering outlined in \cite{Baym}, we consider the following first order self-energy diagrams, generated by the quartic terms in \cref{magnonH}.

\onecolumngrid

\begin{figure}[tph]
\centering
\begin{tikzpicture}[scale=1.1]
\node at (-1,0.2) {$\Sigma_{4}=$};
    \draw[decoration={markings,mark=at position 0.6 with {\arrow{latex}}},postaction=decorate] (0,-1) -- (0,-0.4) node[pos=0.5,left] {$\bold{p}$};
    \draw[decoration={markings,mark=at position 0.6 with {\arrow{latex}}},postaction=decorate] (0,-0.4) -- (0,1) node[pos=0.5,left] {$\bold{k_1}$};
    \draw[decoration={markings,mark=at position 0.6 with {\arrow{latex}}},postaction=decorate] (0,1) -- (0,1.6) node[pos=0.5,left] {$\bold{p}$};
    \draw[dashed] (0,1) -- (1,1);
    \draw[dashed] (0,-0.4) -- (1,-0.4);
    
    \draw[decoration={markings,mark=at position 0.55 with {\arrow{latex}}},postaction=decorate] (1,1) arc (45:-45:1) node[pos=0.5,right] {$\bold{k_3}$};
    \draw[decoration={markings,mark=at position 0.55 with {\arrow{latex}}},postaction=decorate] (1,-0.4) arc (225:135:1) node[pos=0.5,left] {$\bold{k_2}$};

    \node at (2.25,0.2) {$+$};

    \draw[decoration={markings,mark=at position 0.6 with {\arrow{latex}}},postaction=decorate] (3,-1) -- (3,-0.4) node[pos=0.5,left] {$\bold{p}$};
    \draw[decoration={markings,mark=at position 0.6 with {\arrow{latex}}},postaction=decorate] (3,1) -- (3,1.6) node[pos=0.5,left] {$\bold{p}$};
    \draw[dashed] (3,1) -- (5,1);
    \draw[dashed] (3,-0.4) -- (5,-0.4);
    
    \draw[decoration={markings,mark=at position 0.35 with {\arrow{latex}}},postaction=decorate] (3,-0.4) -- (5,1) node[pos=0.3,below] {$\bold{k_1}$};
    \draw[decoration={markings,mark=at position 0.75 with {\arrow{latex}}},postaction=decorate] (5,-0.4) -- (3,1) node[pos=0.75,above] {$\bold{k_2}$};
    \draw[decoration={markings,mark=at position 0.55 with {\arrow{latex}}},postaction=decorate] (5,1) -- (5,-0.4) node[pos=0.5,left] {$\bold{k_3}$};

    \node at (5.5,0.2) {$+$};

    \draw[decoration={markings,mark=at position 0.6 with {\arrow{latex}}},postaction=decorate] (6.5,-1) -- (6.5,-0.4) node[pos=0.5,left] {$\bold{p}$};
    \draw[decoration={markings,mark=at position 0.6 with {\arrow{latex}}},postaction=decorate] (8.5,1) -- (8.5,1.6) node[pos=0.5,right] {$\bold{p}$};
    \draw[dashed] (6.5,1) -- (8.5,1);
    \draw[dashed] (6.5,-0.4) -- (8.5,-0.4);
     \draw[decoration={markings,mark=at position 0.5 with {\arrow{latex}}},postaction=decorate] (6.5,-0.4) -- (6.5,1) node[pos=0.5,left] {$\bold{k_1}$};
    \draw[decoration={markings,mark=at position 0.5 with {\arrow{latex}}},postaction=decorate] (6.5,1) -- (8.5,-0.4) node[pos=0.5,below] {$\bold{k_3}$};
    \draw[decoration={markings,mark=at position 0.55 with {\arrow{latex}}},postaction=decorate] (8.5,-0.4) -- (8.5,1) node[pos=0.5,right] {$\bold{k_2}$};

    \node[anchor=west] at (9,0.2) {$+$ time reversal};

\end{tikzpicture}

\caption{\label{Sigma4} All single-loop order corrections to the Self-Energy generated by the quartic terms in \cref{magnonH}. Following \cite{Baym}, the dashed lines represent vertices given by \cref{V4} and solid lines are magnon-lines. We include among these corrections the time-reversal of each diagram, in order to account for detailed balance \cite{Harris,HarrisAFM}.}
\end{figure}
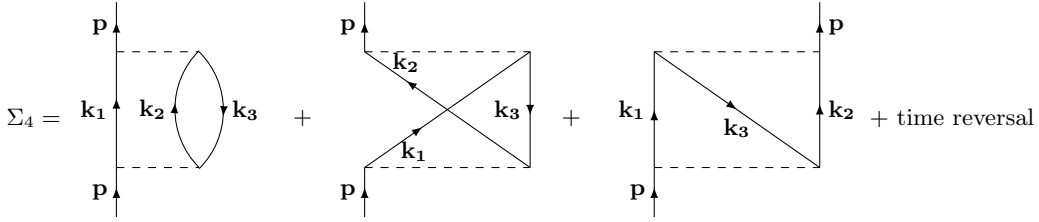

\twocolumngrid
Excluding the time-reversed contributions, the self-energy correction is
\begin{align}
\begin{aligned}
    \Sigma_4=\frac{1}{N^2}\SUM{k_1,k_2,k_3}{}&V^{(4)}_{1,2,p,3}\of{2V^{(4)}_{p,3,1,2}+V^{(4)}_{p,3,2,1}}\times \\ &\times\frac{(1+n(\omega_{k_1}))(1+n(\omega_{k_2}))n(\omega_{k_3})}{i\omega-\omega_{k_1}-\omega_{k_2}+\omega_{k_3}}.
    \end{aligned} \\ \nonumber
\end{align}
The time-reversed diagrams are a consequence of detailed balance, and introduce a factor of $\of{1-e^{-\beta\omega_p}}$ to the scattering width $\Gamma_C=-2\textrm{Im}\of{\Sigma_4}$ \cite{Harris,HarrisAFM}. The width is most easily analyzed in the 'center of momentum' frame. Defining $\bold{K}=(\bold{k_1}+\bold{k_2})/2$ and $\bold{Q}=(\bold{k_1}-\bold{k_2})/2$ and assuming small momenta as in the previous section the width can be approximated roughly as
\begin{widetext}
\begin{align}
    &\Gamma_C\approx \frac{\omega_p}{2\pi\beta^2}\int d^2\bold{K}d^2\bold{Q}\frac{W(\bold{p},\bold{K},\bold{Q})}{\omega_{K+Q}\omega_{K-Q}(2\omega_{K}+2\omega_Q-\omega_p-2\Delta)}\delta\of{2\omega_0\of{p^2+K^2-Q^2-2\bold{p}\cdot\bold{K}}}\eqtxt{,}{where} \\
   &W(\bold{p},\bold{K},\bold{Q})=V^{(4)}_{\bold{K}+\bold{Q},\bold{K}-\bold{Q},\bold{p},2\bold{K}-\bold{p}}\of{2V^{(4)}_{\p,2\K-\p,\K+\Q,\K-\Q}+V^{(4)}_{\p,2\K-\p,\K-\Q,\K+\Q}}
\end{align}
\end{widetext}
In contrast to the previous section, all states are available for this kind of scattering, $\Gamma_C>0$ for all $\bold{p}$. The $\delta-$function is used to eliminate the angular integral for $\bold{K}$. In terms of the dimensionless variable $z$ the width can be written
\begin{align}\label{gc}
    \Gamma_C=\of{\frac{\alpha}{8S}}^2\frac{z+\beta\Delta}{4\beta^3\omega_0^4}I(z,\beta\Delta)
\end{align}
where $I$ is a dimensionless integral given by the rather complicated expression
\begin{widetext}
    \begin{align}
        \label{I}&I(z,\beta\Delta)=\int_0^{z_c}dx\int_{(\sqrt{x}-\sqrt{z})^2}^{(\sqrt{x}+\sqrt{z})^2}dy\frac{F(x,y,z,\beta\Delta)}{\sqrt{\left[\of{\sqrt{x}+\sqrt{z}}^2-y\right]\left[y-\of{\sqrt{x}-\sqrt{z}}^2\right]}}\frac{1}{2(x+y)-z+\beta\Delta}\\
        \nonumber\\
        &F=\frac{1}{a\sqrt{a^2-b^2}}\left[V_1+\frac{a}{b^2}\of{a-\sqrt{a^2-b^2}}V_2+\frac{a\sqrt{a^2-b^2}-a^2+b^2}{b^2}V_3+\frac{\of{a-\sqrt{a^2-b^2}}^2}{b^2}V_4\right],\nonumber\\
        \nonumber\\
        &a=x+y+\beta\Delta,\hspace{1cm}b=\sqrt{2xy},\hspace{1cm}V_1=\frac{1}{4}\of{3x+z-y}{x+z-3y}, \hspace{1cm} V_2=3xy-8\sqrt{xz}y\cos\phi,\nonumber\\
        \nonumber\\
        &V_3=4yz, \hspace{1cm} V_4=4yz\cos^2\phi,\hspace{1cm} \cos\phi=\frac{x+z-y}{2\sqrt{xz}}.\nonumber
    \end{align}
\end{widetext}

\begin{figure}[tph]
\centering
\includegraphics[width=0.5\columnwidth]{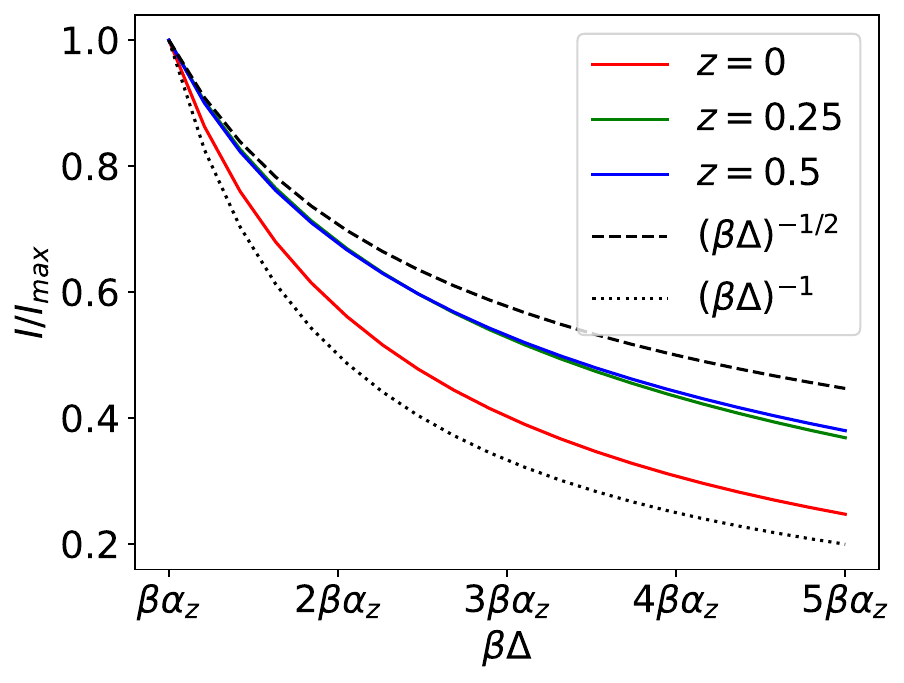}
\caption{\label{Ivbd}Scaling of the dimensionless integral $I$ with parameter $\beta\Delta$.}
\end{figure}

A normalized plot of $I$ vs $\beta\Delta$ is shown in \cref{Ivbd} for several values of $z$. For $z<0.5$, the integral scales as $1/\beta\Delta<I<1/\sqrt{\beta\Delta}$. From this we conclude that the scattering width \cref{gc} scales weakly with the field at small momenta, where the Bose distribution gives the largest contribution. Changes in the magnon resistivity due to conservative scattering are, therefore, governed primarily by the change in magnon number. Because $\Gamma_C$ is only weakly suppressed by the external field, the magnon resistivity, $\rho\sim\Gamma/N_0$, is still enhanced due to the suppression in magnon number. At higher temperatures $\Gamma\approx\Gamma_C$, and so the MMR effect will be much smaller, and governed by changes in magnon number.

\bibliography{biblio}

\end{document}